\documentclass[onecollarge,natbib]{svjour2}
\bibpunct{[}{]}{;}{n}{}{,} % to get "[numbered]" references from natbib
\smartqed  % flush right qed marks, e.g. at end of proof
\usepackage{graphicx}
\usepackage{amssymb,amsmath,epsfig,bm,pifont}
\usepackage{color}

   \newcommand{\RedTn}[1]{\textcolor{red}{#1}}
\newcommand{\Ds}{\displaystyle}                                           %%%%%%%%%
\journalname{Few-Body Systems}
\begin{document}
%\preprint{\hbox{RUB-TPII-04/2014}
\title{On the pion distribution amplitude. %\thanks{Grants or other notes
}
\subtitle{Derivation, properties, predictions}

%\titlerunning{$\pi$ distribution amplitude}        % if too long for running head

\author{N. G. Stefanis \and
        S. V. Mikhailov \and
        A. V. Pimikov
        }

%\authorrunning{Stefanis et al.} % if too long for running head

\institute{N. G. Stefanis \at
           Institut f\"{u}r Theoretische Physik II,
           Ruhr-Universit\"{a}t Bochum,
           D-44780 Bochum, Germany \\
           %Tel.: +49-234-3223724\\
           %Fax: +49-234-3214697\\
           \email{stefanis@tp2.ruhr-uni-bochum.de}           %  \\
           \and
           S. V. Mikhailov \at
           Bogoliubov Laboratory of Theoretical Physics,
           JINR, 141980 Dubna, Russia\\
           \email{mikhs@theor.jinr.ru}
           \and
           A. V. Pimikov \at
           Bogoliubov Laboratory of Theoretical Physics,
           JINR, 141980 Dubna, Russia\\
           \email{pimikov@theor.jinr.ru}
}

\date{Received: date / Accepted: date}

\maketitle

\begin{abstract}
We provide an in-depth analysis of the $\pi$ distribution
amplitude in terms of two different Gegenbauer representations.
Detailed predictions for the $\pi-\gamma$ transition form factor
are presented, obtained with light-cone sum rules.
Various $\pi$ distribution amplitudes are tested and the
crucial role of their endpoint behavior in the form-factor analysis
is discussed.
Comparison with the data is given.

%Include up to five keywords.
\end{abstract}

\section{Introduction}
\label{intro}
The pion --- the lightest meson --- epitomizes the bound state of two
light quarks within Quantum Chromodynamics (QCD).
Employing light-front (LF) quantization, recently reviewed in
\cite{Bakker:2013cea}, the $n$-parton wave function of the pion is
defined in a frame-independent way by
$\psi_{n/\pi}(x_{i}, \bm{k}_{\perp i}, \lambda_{i})$.
It is advantageous to consider instead, in lieu with QCD collinear
factorization, the LF distribution amplitude (DA) \cite{LB80}, which
is the LF wave function integrated over transverse momentum at fixed
longitudinal momentum fraction
$x_i=(k^0+k^3)/(P^0+P^3)=k^+/P^+$
of the pion's momentum $P$, with $\sum_{i=1}^{n}x_i=1$.
Then, the leading-twist-two DA of the $\bar{q}q$ valence state is
described by
$
  \varphi_{\pi}^{(2)} \left(x,\mu^2\right)
=
  \int_{}^{\mu^{2}} d\bm{k}_{\perp}^{2}(16\pi^2)^{-2}
  \psi\left( x, \bm{k}_{\perp}\right)
$.
In operator language, this $\pi$ DA is given by the following
gauge-invariant matrix element:
\begin{equation}
  \langle 0| \bar{q}(z) \gamma_\mu\gamma_5 [z,0] q(0)
           | \pi(P)
  \rangle|_{z^{2}=0}
=
  if_\pi P_\mu \int_{0}^{1} dx e^{i x (z\cdot P)}
  \varphi_{\pi}^{(2)} \left(x,\mu^2\right) \, ,
\label{eq:pion-DA}
\end{equation}
%Eq (1)
where the gauge link
$
 [z,0]
=
 \mathcal{P}\exp \left(
                       ig \int_{0}^{z} A^{\mu}d\tau_{\mu}
                 \right)
 =1
$
by virtue of imposing the lightcone gauge $A^+=0$.
The momentum-scale dependence of the DA is controlled within
perturbative QCD by a renormalization-group type evolution equation
due to Brodsky, Lepage \cite{LB80}, and
Efremov, Radyushkin \cite{Efremov:1978rn} (ERBL).
For this reason, one seeks to express the pion DA in terms of the
eigenfunctions of this equation, i.e., the Gegenbauer harmonics
$\psi_n(x)=6x(1-x) C^{(3/2)}_n(2x-1)$
with the Gegenbauer polynomials $C^{(3/2)}_n(\xi=2x-1)$:
\begin{equation}
  \varphi_{\pi}^{(2)}(x,\mu^2)
=\psi_0(x)
  + \sum_{n=2,4, \ldots}^{\infty} a_{n}(\mu^2) \psi_n(x)
  \, ,
\label{eq:gegen-exp}
\end{equation}
%Eq (2)
where
$\psi_0(x)=\varphi_{\pi}^{\rm asy}(x)=6x(1-x)\equiv 6x\bar{x}$
is the asymptotic DA and the nonperturbative content of the pion
bound valence state is encoded in the expansion coefficients
$a_{n}(\mu^2)$.
Owing to the fact that the $\pi$ DA (or any other hadron DA) is not
directly measurable, one has to model it by determining the
coefficients $a_{n}(\mu^2)$ at a typical hadronic scale of order
$\mu^2\sim 1$~GeV$^2$ using some nonperturbative method
(see next section) and use ERBL evolution to obtain them
at any scale $Q^2$.
Regardless of the differences among these methods, the pion DAs share
some particular properties which will be addressed in
Sec. \ref{sec:pi-prop}.
In fact, the success or failure of a particular DA will depend on the
ability to describe existing data for various pion observables and make
tangible predictions for data to come (more in Sec. \ref{sec:predict}).
Our findings will be summarized in Sec. \ref{sec:concl}, where we will
also draw our conclusions.

\section{Derivation of the $\pi$ DA}
\label{sec:pi-DA}
\begin{figure}[t]
\centering
\includegraphics[width=0.45\textwidth]{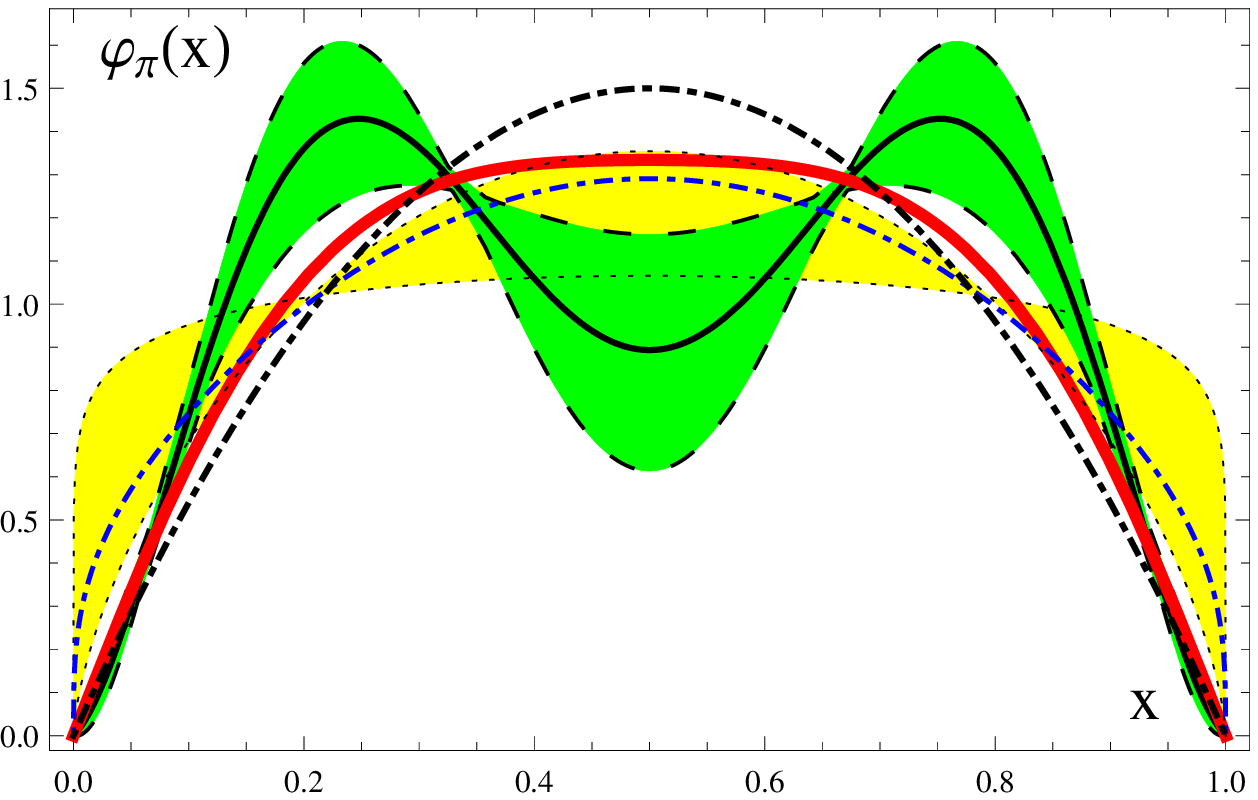}  %BMS.DSE.plat.4.eps
\hfill
\includegraphics[width=0.45\textwidth]{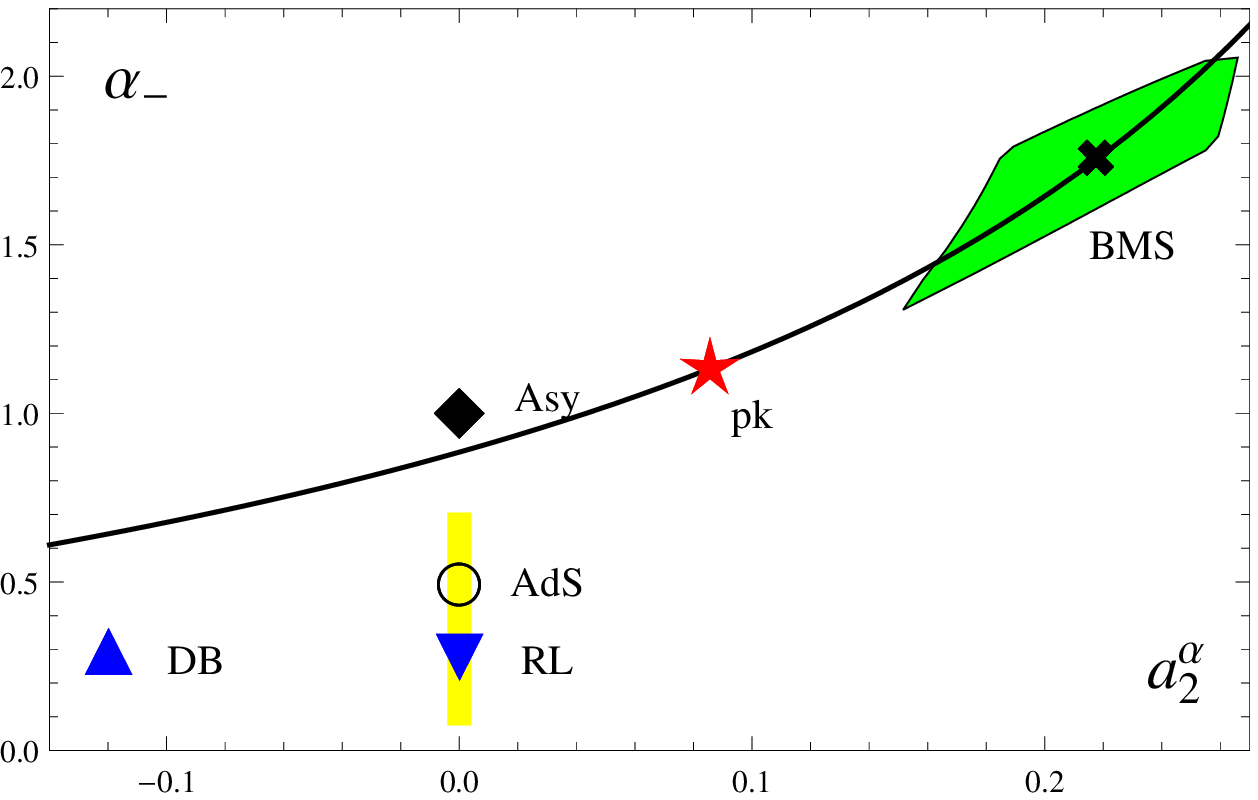}  %fig-a2.alf.DAs.inverseMomLine.3.13.eps
\caption{
\textit{Left panel}: Pion DAs obtained within our NLC-SR method
(dark-shaded green area), with the solid line inside it
depicting the BMS DA \protect\cite{Bakulev:2001pa}.
The light-shaded yellow region of broad DAs shows the DSE results,
the dashed-dotted line representing the DB DA
\protect\cite{Chang:2013pq}.
The thick red line denotes the new platykurtic DA obtained in
\protect\cite{MPS14}
and presented in \protect\cite{Stefanis:2014nla}.
\textit{Right panel}:
Pion DAs plotted in the ($\alpha_-,a_{2}^{\alpha}$) space.
The curve denotes the orbit of DAs with the same value of
$\langle x^{-1}\rangle_\pi \approx 3.13$.
The normalization scale for both panels is $\mu^2=$4 GeV$^2$.}
\label{fig:pi-DAs}
\end{figure}

The pion DAs used in our analysis were determined with the help of QCD
sum rules with nonlocal condensates (NLC-SR for short)
\cite{Bakulev:2001pa}.
Our method derives from \cite{MR86} in which the meson DA
$\varphi_\text{M}(x)$ over the longitudinal momentum fractions
$x \cdot P$ was linked to the distributions $\Phi_j$ of the virtuality
of quarks and gluons in the nonperturbative vacuum of QCD.
This was achieved by relating the pion and its first resonance $A_1$ to
the correlator of two axial currents with NLC contributions in a sum
rule:
\begin{eqnarray}
  f_{\pi}^2\varphi_\pi(x) +
  f_{A_1}^2\varphi_{A_1}(x)\exp\left(-\frac{m^2_{A_1}}{M^2}\right)
&=&
   \int_{0}^{s_{\pi}^0}\rho^{\rm pert}_{\rm NLO}(x;s)e^{-s/M^2}ds
  +
   \frac{\langle \alpha_{\rm s} GG \rangle}{24\pi M^2}\
   \Phi_{\rm G}\left(x;M^2;\Delta\right) \nonumber \\
&+&  \frac{8\pi\alpha_{\rm s}\langle{\bar{q}q\rangle}^2}{81M^4}
      \sum_{i={\rm S,V,T_{1,2,3}}}\Phi_i\left(x;M^2;\Delta\right) \, .
\label{eq:nlcsrda}
\end{eqnarray}
%Eq (3)
Here the index $i$ runs over scalar (S), vector (V) quark condensates,
and quark-gluon condensates (T), while (G) denotes the gluon
condensate.
The most important parameter is the nonlocality
$\Delta=\lambda_{q}^{2}/(2M^2)$,
where $M^2$ is the Borel parameter and $\lambda_{q}^{2}$ is the average
vacuum-quark virtuality, defined by
$
  \lambda_{q}^{2}
=
  \langle
  \bar{q}igG^{\mu\nu}\sigma_{\mu\nu}q \rangle / 2\langle
        \bar{q}(0)q(0)
  \rangle \approx [0.35-0.45]$~GeV$^2
$.
This sum rule depends on the duality interval $s_{\pi}^0$ in the axial
channel and involves the spectral density
$\rho^{\rm pert}_{\rm NLO}(x;s)$ in next-to-leading order (NLO)
of perturbative QCD \cite{Bakulev:2001pa,BM98}.
A whole family of admissible $\pi$ DAs can be reconstructed
\cite{Bakulev:2001pa,Stefanis:2012yw} from the above sum rule via the
calculation of the moments $ \langle \xi^N \rangle$ and also in terms
of an independent sum rule for the inverse moment
$\langle x^{-1}\rangle_\pi$:
\begin{equation}
  \langle \xi^N \rangle
=
 \int_{0}^{1}dx \varphi_{\pi}^{(2)}(x, \mu^2)(2x-1)^N;~~
 \langle x^{-1}\rangle_\pi
=
 \int_0^1 \frac{\varphi_{\pi}(x)}{x}dx \, .
\label{eq:moments}
\end{equation}
%Eq (4)
Below, we only summarize the main features of NLC-SRs, referring the
reader for further details to \cite{MR86,Bakulev:2001pa,BMS04kg}.
(i) We start with the same condensates as in the standard (local)
approach \cite{Chernyak:1983ej} but keep the nonlocal quantities
$\langle \bar{q}(0)\Gamma q(z) \rangle$,
$\langle G_{\mu \nu}(0)G_{\mu \nu}(z) \rangle$, etc.
unexpanded.
In this way, we obtain in Eq. (\ref{eq:nlcsrda}) 6 functions $\Phi_j$
instead of 2 numbers for
$\langle \bar{q}(0)\Gamma q(0) \rangle,
\langle G_{\mu \nu}(0)G_{\mu \nu}(0) \rangle$.
(ii) The nonlocal condensates are assumed to decrease with increasing
distance (between vacuum fields).
In fact, the \textit{finiteness} of the widths of the NLCs plays a
crucial role because it renders the RHS of Eq. (\ref{eq:nlcsrda}) less
singular in $x$ \cite{MR86}.
To be specific, the singularities at the end points $x=0,1$,
inevitable in the local approach, disappear.
These finite widths have been estimated from the next higher terms
in the expansions of the NLCs and are proportional to
$\sim 1/\lambda_q^2$.
(iii) To estimate the integral characteristics of $\varphi_\pi(x)$
in Eq. (\ref{eq:moments}), it is plausible to approximate the
NLC by Gaussian models which depend only on the single scale
$\lambda_q^2/2$.
This is sufficient to account for the main effect,
i.e., the finite widths of the NLCs.
The complete calculation of all NLO corrections to the RHS of the SR
above will be undertaken in a separate work \cite{MPS14}.
(iv) The evaluation  of the SR in (\ref{eq:nlcsrda}) enables us to
estimate the first 5 moments
$\langle \xi^{2,4,6,8,10} \rangle_\pi$ and, independently, the value of
$\langle x^{-1}\rangle_\pi$.
Making use of these estimates, a set of admissible pion DAs,
was determined in terms of 2 Gegenbauer harmonics
\cite{Bakulev:2001pa,BMS04kg}, which we display on the LHS of
Fig. \ref{fig:pi-DAs} in terms of a dark-shaded green area.
The central curve within this band is termed the ``BMS model''
\cite{Bakulev:2001pa}.
At the scale $\mu^2 \approx 1$~GeV$^2$ it is given by
$\{a_2^\text{BMS}\approx 0.2,a_4^\text{BMS}\approx -0.14\}$
using expansion (\ref{eq:gegen-exp}).

%%%%%%%%%%%%%%%%%%%%%%%%%%%%%%%%%%%%%%%%%%%%%%%%%%%%%%%%%%%%%%%%%%%%%%% Figure 1

Let us now turn our attention to another powerful method to extract the
pion DA which is based on QCD's Dyson-Schwinger equations (DSE) ---
see \cite{Boucaud:2011ug} for a general review and \cite{Cloet:2013jya}
for a review pertinent to the calculation of parton distribution
amplitudes and functions for hadrons.
The DSE-based derivation of the pion DA employs the determination
of its first 50 $\langle x^n\rangle$ moments from which the pion DA is
reconstructed in the form of the following Gegenbauer representation:
\begin{equation}
  \varphi_{\pi}(x, \mu^2)
=
  f(\{\alpha,a^\alpha_2,...,a^\alpha_{j_s}\},x)
=
  \psi^{(\alpha)}_0(x)+ \sum_{j=2,4,\ldots}^{j_{s}}
        a_{j}^{\alpha}(\mu^2)\psi^{(\alpha)}_n(x)
\label{eq:DA-gen-Gegen}
\end{equation}
%Eq (5)
with
$
 \psi^{(\alpha)}_n(x)
=
 N_{\alpha}(x\bar{x})^{\alpha_{-}}C^{(\alpha)}_{n}(2x-1)
$,
and
$N_{\alpha}=1/B(\alpha + 1/2, \alpha + 1/2)$,
$\alpha_{-}=\alpha -1/2$.
Here $\alpha$ is considered to be a fit parameter and is not
forced to the value 3/2, i.e., to the conformal expansion.
Two different procedures to treat the gap and the
Bethe-Salpeter kernels in incorporating the nonperturbative features
of dynamical chiral symmetry breaking (DCSB) were used in
\cite{Chang:2013pq,Cloet:2013tta}, based on the rainbow-ladder (RL)
truncation and the DCSB-improved (DB) kernel.
The corresponding pion DAs are described by
$
 \varphi_{\pi}^{\rm RL}(x)
=
 1.74 (x\bar{x})^{\alpha_{-}^{\rm RL}}
 [1 + a_{2}^{\rm RL} C_{2}^{(\alpha^{\rm RL})}(x-\bar{x})]
$
with
$\alpha_{-}^{\rm RL}=0.29$, $a_{2}^{\rm RL}=0.0029$
and
$
 \varphi_{\pi}^{\rm DB}(x)
=
 1.81 (x\bar{x})^{\alpha_{-}^{\rm DB}}
 [1 + a_{2}^{\rm DB} C_{2}^{(\alpha^{\rm DB})}(x-\bar{x})]
$
with
$\alpha_{-}^{\rm DB}=0.31$, $a_{2}^{\rm DB}=-0.12$.
The profiles of the DAs, determined in \cite{Cloet:2013tta} at the
scale $\mu^2=4$~GeV$^2$, are shown on the left panel of
Fig. \ref{fig:pi-DAs} in the form of the lower light-shaded yellow
area within the range of errors on
$\alpha_{-}=0.35^{+0.32}_{-0.24}$ and $a_2^\alpha=0$.
This domain of $\alpha_{-}$ and $a_2^{\alpha}$ values incidentally
includes the unrelated AdS/QCD model of \cite{Brodsky:2011yv} with
$\psi^{(1)}_0(x)$ and $\alpha_-=1/2$,
depicted by an open circle on the right panel of Fig. \ref{fig:pi-DAs}.
In order to compare our DAs with this sort of DAs, we switch from the
conformal expansion (\ref{eq:gegen-exp})
to the general expansion (\ref{eq:DA-gen-Gegen}),
employing the approximation of our DAs in terms of the lowest two
terms  $a_2$ and $a_4$, viz.,
\begin{eqnarray}
  a_2
=
  \frac{7}{18} \int\limits_0^1\!\! dx
  f(\{\alpha,a_2^\alpha\},x) C_2^{(3/2)}(2x-1) \, ; ~~~
  a_4
=
  \frac{11}{45}\int\limits_0^1\!\! dx
  f(\{\alpha,a_2^\alpha\},x) C_4^{(3/2)}(2x-1) \, .
\label{eq:a2a4tob2alf}
\end{eqnarray}
%Eq (6)
Solving this set of equations with respect to $(\alpha, a_2^\alpha)$,
we get an approximate representation for each member of the BMS
set of DAs, defined by $(a_2,a_4)$, in terms of $(\alpha, a_2^\alpha)$
within the expansion Eq. (\ref{eq:DA-gen-Gegen}) for $j_s=2$.
The result of this mapping for the whole BMS set of DAs is graphed on
the right panel of Fig. \ref{fig:pi-DAs} as a dark-shaded green area.
The other depicted DAs are identified appropriately in the figure.

\section{Properties of the $\pi$ DA}
\label{sec:pi-prop}
As one sees from the left panel of Fig. \ref{fig:pi-DAs}, the profiles
of the pion DA can vary significantly.
Lattice simulations, while useful, can currently provide constraints
only for the second moment $\langle \xi^2 \rangle$
(or equivalently the Gegenbauer coefficient $a_2$)
\cite{Braun:2006dg,Arthur:2010xf}, while higher moments (coefficients)
are still at large.
Thus, lacking experimental evidence to restrict this variation, one may
appeal to other \textit{unbiased} tools to explore the structure of the
pion DA.
To this end, one of us has recently proposed \cite{Stefanis:2014nla} to
use the Kuramoto model (see \cite{Strogatz2000} for a review) which
describes the synchronization of nonlinear oscillators in complex
systems.
The main features of Stefanis's approach are as follows.
The longitudinal momentum fractions $x$ are considered to be the
natural frequencies (phases) of a large number ($N\to\infty$) of
phase-coupled oscillators.
The interaction among these oscillators gives rise to characteristic
patterns of clustering in the interval $x\in[0,1]$ which in turn
correspond to particular DA profiles.
The analysis in \cite{Stefanis:2014nla} has revealed that at finite
momentum values of $Q^2$, the $x$ spectrum tends to synchronize as a
result of nonperturbative correlations, encoded in nonlocal
condensates, and the DCSB which is responsible for the
generation of quark and gluon masses within a DSE-based framework.
While the first effect suppresses the endpoints $x=0,1$, the DSE
treatment yields strong enhancement of these regions.
This contradicting behavior inspired in \cite{Stefanis:2014nla}
the following parametrization of the pion DA
$
 \varphi_{\pi}^\text{``true''}(x)
\approx
 a\varphi_{\pi}^\text{BMS}(x) + (1-a)\varphi_{\pi}^\text{DSE}(x)
$
with a mixing parameter $a\approx 0.7 - 0.9$.

Motivated by this finding, we furthered our NLC-SR-based approach and
selected an admissible DA \cite{MPS14} that inherently combines the
synchronization properties mentioned above and detailed in
\cite{Stefanis:2014nla}.
The profile of this DA, shown at the scale of $4$~GeV$^2$ as a thick
red solid line on the LHS of Fig. \ref{fig:pi-DAs}, is characterized
by a broad maximum along a downward concave curve that bears a strong
resemblance to the DSE DAs (DB and RL) over a large range of $x$
values.
However, its tails at $x=0$ and $x=1$ are suppressed, so that this DA
has a platykurtic (pk) profile, in contrast to the family of the
broad, endpoint-enhanced DSE DAs (light-shaded yellow band bounded
by dotted lines on the LHS of Fig. \ref{fig:pi-DAs}).
At the considered scale $\mu^2=4$~GeV$^2$ it is given by
$a_{2}^{\rm pk}\approx 0.057$ and $a_{4}^{\rm pk} \approx-0.013$
in terms of the conformal expansion in Eq. (\ref{eq:gegen-exp}).
Its position is marked on the right panel of this figure by the symbol
\RedTn{$\bigstar$} at
$(\alpha_{-}\approx 1.16, a_2^{\alpha}\approx 0.09)$.
At the midpoint it gives
$\varphi_{\pi}^{\rm pk}(x=1/2, \mu^2=4~{\rm GeV}^2)=1.33$,
which is close to
$\varphi_{\pi}^{\rm RL}(x=1/2, \mu^2=4~{\rm GeV}^2)=1.16$
and
$\varphi_{\pi}^{\rm DB}(x=1/2, \mu^2=4~{\rm GeV}^2)=1.29$
\cite{Chang:2013pq}
and conforms well with the sum-rule estimate
$\varphi_{\pi}(x=1/2)=1.2 \pm 0.3$
computed in \cite{BF89}.
Thus, the platykurtic DA can be interpreted as a realization of
the Kuramoto-inspired parametrization proposed in
\cite{Stefanis:2014nla}.
As we will see in the next section, it leads to a prediction
for the pion-photon TFF which is very close to that obtained
with the original BMS DA.

\section{Pion-photon transition and other $\pi$ observables}
\label{sec:predict}
\begin{figure}[ht!]
\centering
\includegraphics[width=0.50\textwidth]{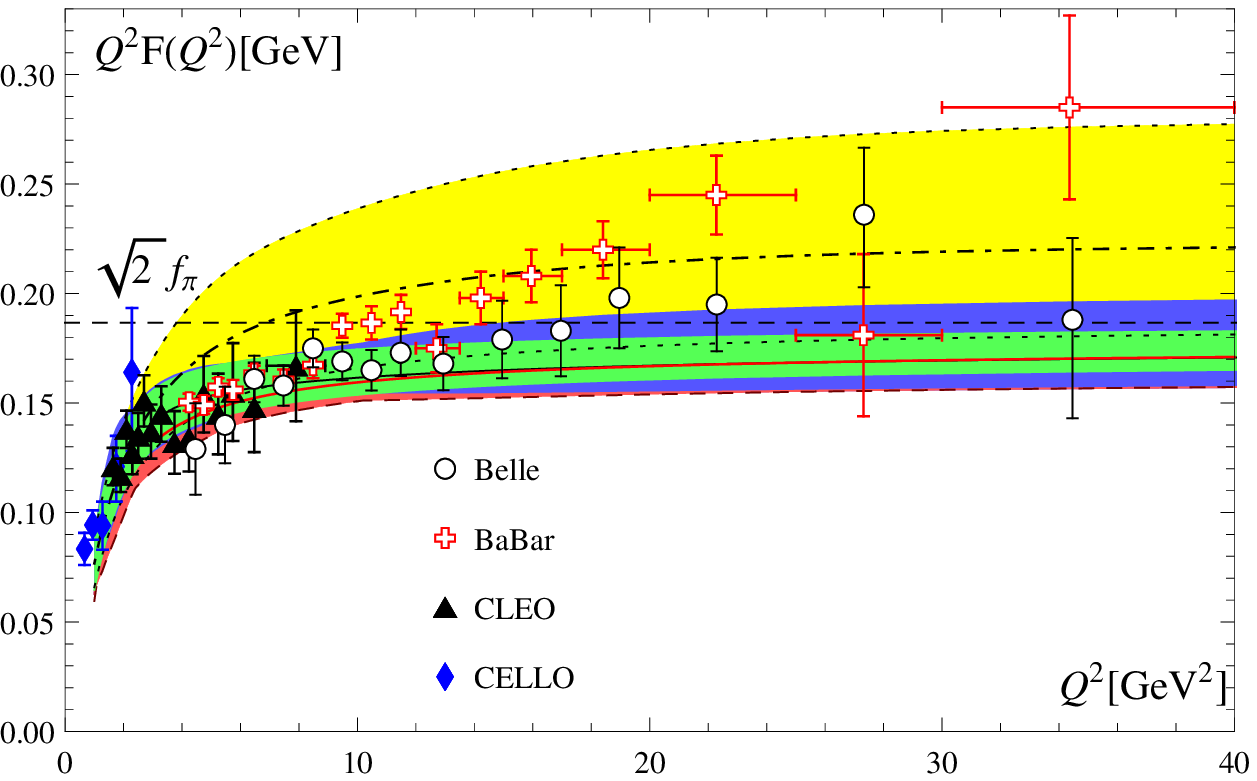}   %BMS.3DvsData.DSE.plat.eps
\hfill
\includegraphics[width=0.43\textwidth]{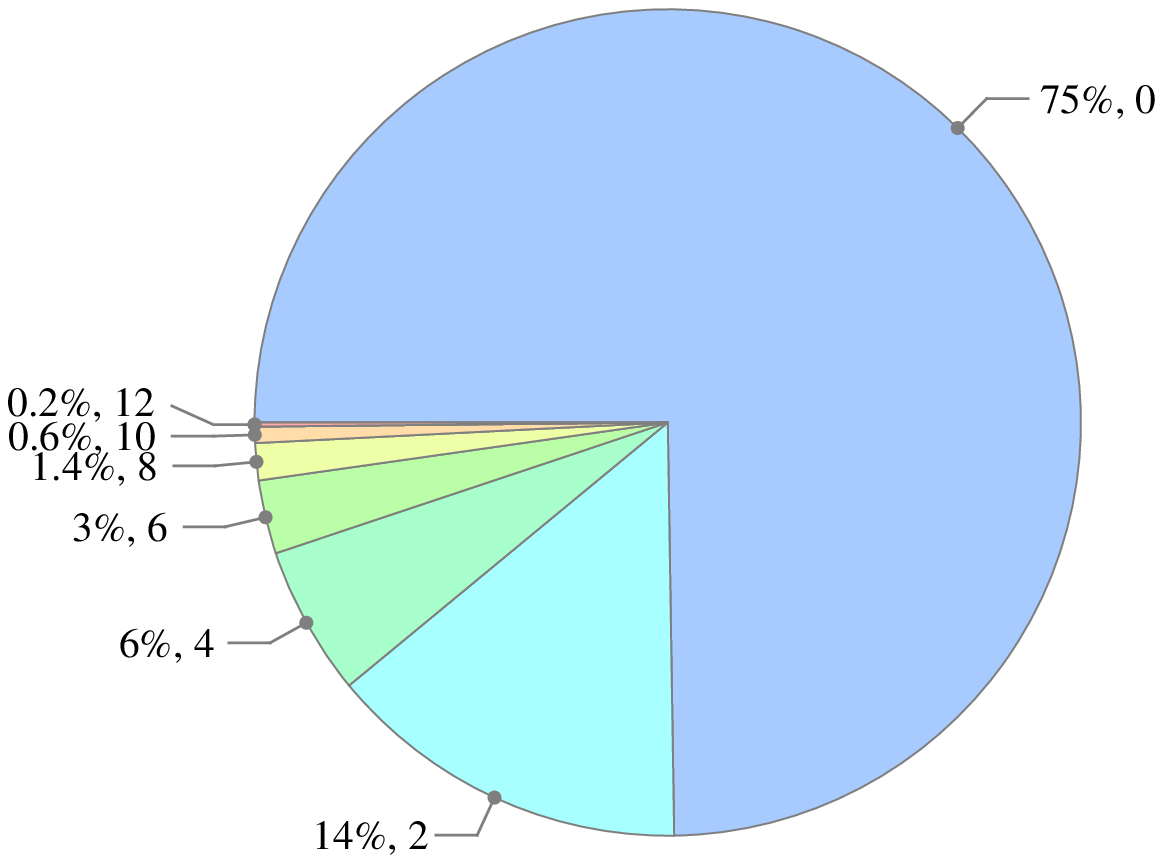}
\caption{
\textit{Left panel}:Predictions for
$Q^2F^{\gamma^*\gamma\pi^0}(Q^2)$ from our LCSR-based approach
using as input BMS DAs \protect\cite{Bakulev:2001pa}
(lower, narrower green band)
in comparison with the existing data and results we obtained with the
DAs of the DSE framework \protect\cite{Cloet:2013tta}
(upper, wider band bounded by dotted lines)
by including the first ten coefficients $a_n$.
The solid (red) line inside the inner shaded band shows the
prediction obtained with the short-tailed platykurtic DA
\protect\cite{MPS14}.
\textit{Right panel}:  Incremental inclusion of contributions to
$Q^2F^{\gamma^*\gamma\pi^0}(Q^2)$ stemming from the coefficients $a_n$
from $n=0$ up to $n=12$
in the representation of the DB DSE DA for
$\alpha_{-}=0.35$ at the scale $Q^2=20$~GeV$^2$.
\label{fig:BMS.3DvsData.DSE.plat}}
\end{figure}

Our predictions for $Q^2F^{\gamma^*\gamma\pi^0}(Q^2)$ vs. $Q^2$
in comparison with the currently existing data
\cite{CELLO91,CLEO98,BaBar09,Belle12}
are shown in Fig. \ref{fig:BMS.3DvsData.DSE.plat} (left panel) in the
form of a dark-shaded green band surrounded by a narrower blue one.
%%%%%%%%%%%%%%%%%%%%%%%%%%%%%%%%%%%%%%%%%%%%%%%%%%%%%%%%%%%%%%%%%%%%%%% Figure 2
The underlying DAs are the bimodal curves, obtained with NLC-SRs
\cite{Bakulev:2001pa}, within the dark-shaded band on the left panel.
One observes that these predictions start to scale with $Q^2$ already
at 10~GeV$^2$.
They are genuine state-of-the-art results because they comprise, within
the framework of lightcone sum rules (LCSR), see, e.g., \cite{Kho99},
\textit{all} contributions to this TFF, currently known within QCD.
These are:
(i) NLO radiative correction,
(ii) twist-four contribution,
(iii) NLO ERBL evolution and
(iv) quantified uncertainties related to the
NNLO \cite{Mikhailov:2009kf} and the
twist-six corrections \cite{Agaev:2010aq},
as we detailed in \cite{Bakulev:2011rp} (inner wider green band),
(v) the effect of a finite virtuality of the untagged quasi-real
photon, included by employing the cut imposed in the Belle experiment
($q_{2}^{2}\approx 0.04$~GeV$^2$,
see \cite{Stefanis:2012yw}) --- narrow red strip below the outer band,
while the effect of the extension of the BMS set to include in
the analysis the $\psi_6(x)$ harmonic, leads \cite{Stefanis:2012yw} to
the outer blue band.
The root cause for the closeness of the calculated curves for the
scaled pion-photon TFF with the bimodal BMS DA and the unimodal
platykurtic DA can be traced to the fact that, for the leading twist
contribution within the conformal expansion, one has
$
 3/(\sqrt{2}f_{\pi})Q^2F_{\gamma^*\gamma\pi^0}^{(\rm LO)}(Q^2)
= \langle x^{-1}\rangle_{\pi}
= 3(1+a_{2}+a_{4} \ldots)
$
\cite{Bakulev:2001pa,Bakulev:2002uc,Bakulev:2003cs},
Hence, DAs with very different coefficients of the conformal expansion
in Eq. (\ref{eq:gegen-exp}) --- and thus shapes --- can nevertheless
yield very close predictions for the pion-photon TFF, provided the
corresponding inverse moment has similar values.
In the case of the general expansion (\ref{eq:DA-gen-Gegen}), one gets
$
 \Ds \langle x^{-1}\rangle_{\pi}
=
 [4\alpha/(2\alpha-1)](1+a_{2}^{\alpha}+\ldots)
$,
so that the same effect appears for those DAs near the hyperbola
shown on the RHS of Fig. \ref{fig:pi-DAs}.

The predictions obtained within our LCSR scheme for the pion-photon
TFF, using as input the DSE DAs from \cite{Cloet:2013tta}, are graphed
on the LHS of Fig. \ref{fig:BMS.3DvsData.DSE.plat} in terms of the
wider upper yellow band within the dotted lines.
Inside this band the prediction due to the DB DA is depicted
explicitly as a dashed-dotted line.
Two observations are worth noting: First, the enlarged BMS band,
which incorporates all mentioned contributions, overlaps with the
DSE one along a narrow strip above $\sim 8$~GeV$^2$.
Second, the DB DA, favored in \cite{Chang:2013pq},
obviously overshoots most experimental data in the whole $Q^2$ range.
This discrepancy becomes more accentuated for the broader RL DA,
obtained in \cite{Chang:2013pq} by imposing a rainbow-ladder truncation
of the Bethe-Salpeter kernels in the DSEs.
In this context we mention that the DA derived within a holographic
AdS/QCD approach \cite{Brodsky:2011yv}, notably,
$\varphi_{\pi}^{\rm AdS/QCD}(x)=(8/\pi)(x\bar{x})^{1/2}$
agrees better with the data relative to the DSE DAs
(cf. the graphics on the RHS of Fig. 2 in
Ref. \cite{Mikhailov:2014rqa}), the reason being that its endpoint
regions are less pronounced.
For similar reasons, the endpoint-concentrated
Chernyak-Zhitnitsky DA \cite{Chernyak:1983ej} overshoots all data as
well, see \cite{Bakulev:2011rp}.

%%%%%%%%%%%%%%%%%%%%%%%%%%%%%%%%%%%%%%%%%%%%%%%%%%%%%%%%%%%%%%%%%%%%%%% Figure 3
\begin{figure}[b!]
\centering
\includegraphics[width=0.45\textwidth]{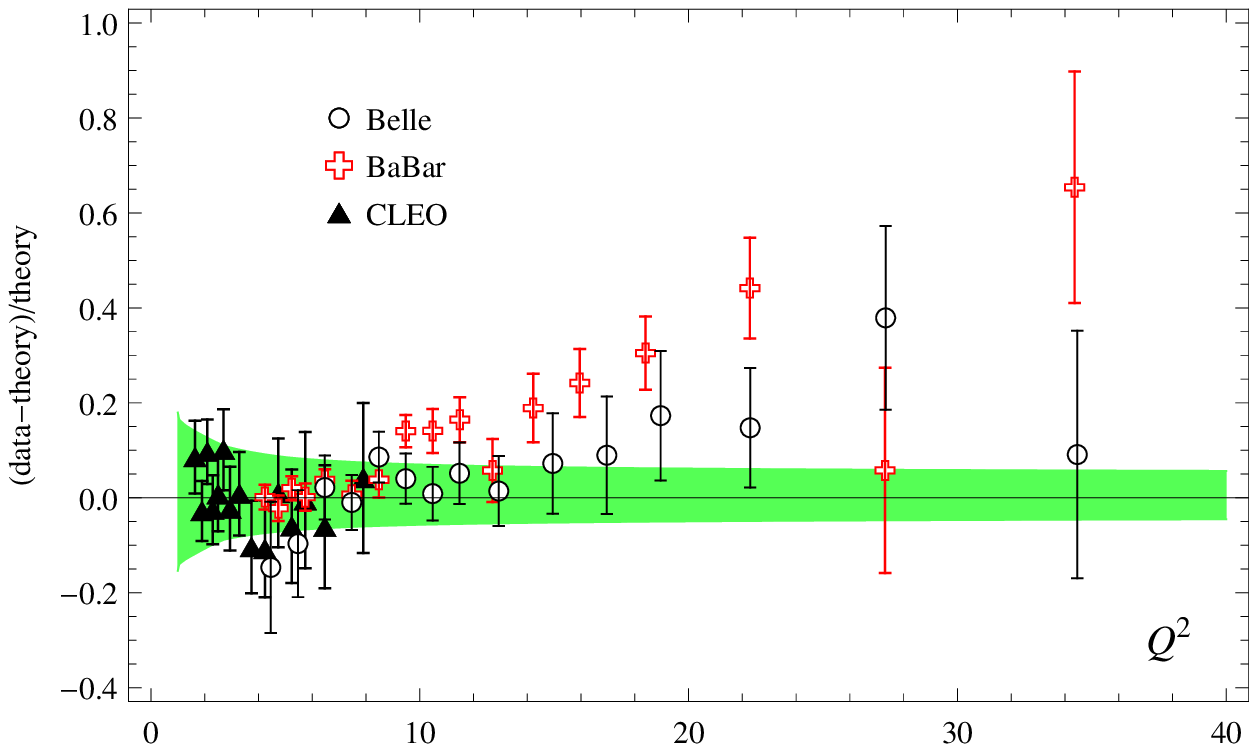}   %fig-Theor.vs.Data.eps
\hfill
\includegraphics[width=0.45\textwidth]{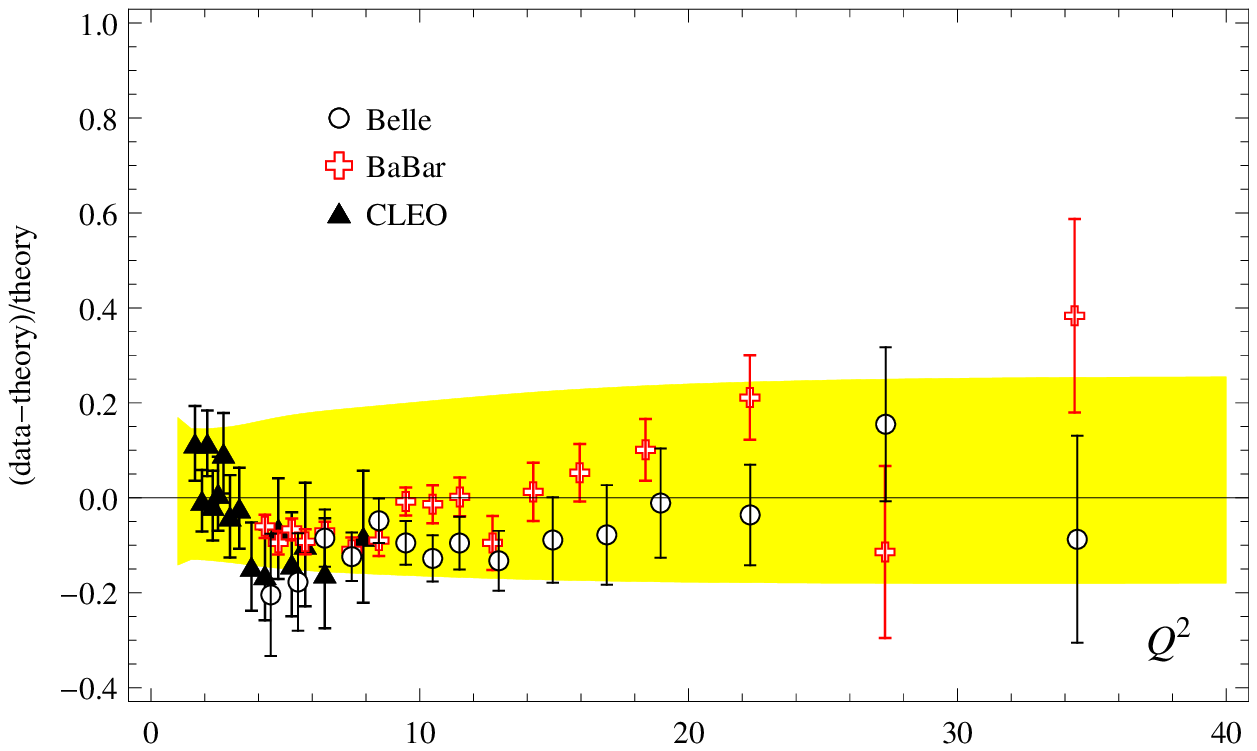}   %fig-Theor.vs.Data.DSE.eps
\caption{(Data-Theory)/Theory vs. $Q^2$.
Theory (dark-shaded green area) is here our LCSR-based approach using
as input the BMS DAs (bimodal and platykurtic) (left panel).
The analogous illustration for the DSE DAs is shown in the right
panel --- wide yellow band.
The computational technique is described in the text.}
\label{fig:data-theory}
\end{figure}

It is important to make some remarks on the computation of the above
results.
It was shown in \cite{Chang:2013pq} that the expansion in
Eq. (\ref{eq:DA-gen-Gegen}) provides a stable result already at
$j_s=2$, with $j_s=4$ producing negligible changes.
On the other hand, in order to include ERBL evolution, one has to go
over to the conformal expansion of Eq. (\ref{eq:gegen-exp}).
The authors of \cite{Chang:2013pq} argued that one has to
include terms of order $n\geq 14$ to ensure a stable conformal
expansion of the pion DA.
Though this may be right for the DSE DA, it is not necessarily true
for the pion TFF calculated with it.
Indeed, we investigated the rate of convergence of the pion-photon TFF,
calculated within our scheme with the DSE DAs, quantitatively
and found that already the contribution to the TFF stemming from the
inclusion of the coefficient $a_{12}$ plays only a marginal role
($0.2 \%$) --- RHS of Fig. \ref{fig:BMS.3DvsData.DSE.plat}.
Thus, our results obtained with the DSE DAs on the LHS of
the same figure are unbiased predictions for
$Q^2F^{\gamma^*\gamma\pi^0}(Q^2)$ of the DSE approach.

To further quantify these statements, we plot in
Fig. \ref{fig:data-theory} the quantity
(Data-Theory)/Theory against $Q^2$, using as ``Theory'' our
LCSR framework in connection with various DAs.
The left panel of this figure shows the result for the bimodal BMS DAs
(dark-shaded green strip), whereas the right panel displays the
analogous graph for the DSE DAs light-shaded yellow band within the
same scheme.
The deviations of the high-$Q^2$ BaBar data from ``Theory'' with the
BMS DAs are larger relative to those with the DSE ones.
This is because the BMS predictions go approximately to a constant at
high $Q^2$ as the lower band in the left plot in
Fig. \ref{fig:BMS.3DvsData.DSE.plat} reveals, while the DSE results
(upper band) do not exclude an auxetic TFF behavior.
In contrast, the platykurtic DA rules out a growth of the TFF and
supports a scaling behavior at higher $Q^2$.
Note that model II, obtained with LCSRs in \cite{Agaev:2010aq} and
retrofitted to the Belle data in the second entry of
\cite{Agaev:2010aq}, lies within the upper blue strip of our
predictions.

Our NLC-SR-based pion DAs can be successfully used to describe
other processes involving the pion DA.
Examples are the pion elastic form factor \cite{Bakulev:2004cu},
the diffractive di-jet production \cite{Bakulev:2003cs}, and the
meson-induced Drell-Yan production for the process
$\pi^-N \rightarrow \mu^+\mu^- X$ \cite{BST07}.
Two major findings of these analyses are worth mentioning here:
(i) Using the convolution scheme developed in \cite{BISS02},
it was shown in \cite{Bakulev:2003cs} that the derived predictions
with the bimodal BMS DAs are in good agreement with the E791 di-jet
events \cite{E79101}--- even in the central $x$ region where the BMS
DA ``bunch'' has its largest uncertainties
(see left panel of Fig. \ref{fig:pi-DAs}).
(ii) It was found in \cite{BST07} that with the BMS DAs the angular
parameters $\lambda,~\mu,~\nu$ to describe the angular distribution of
$\mu^+$ in the unpolarized Drell-Yan $\pi^-N \rightarrow \mu^+\mu^- X$
process comply best with the existing data.

\section{Conclusions}
\label{sec:concl}
We have presented detailed predictions for the pion-photon TFF which
comprise all currently known QCD perturbative and nonperturbative
contributions within a LCSR theoretical scheme.
Within this scheme, we used several types of pion DAs as
nonperturbative input on account of collinear QCD factorization.
We found that BMS-like DAs, which are two-parametric bimodal
distributions \cite{Bakulev:2001pa} replicate those data which are
compatible with scaling, while they disagree with the high-$Q^2$
BaBar data which grow with the momentum.
Crucially, this agreement is tightly connected to the endpoint behavior
of these pion DAs and to much lesser extent to the broadness of the DA
in the central region of $x$.
Indeed, using QCD sum rules with nonlocal condensates, we determined a
new DA which is a short-tailed platykurtic curve but leads to similar
TFF results as the original BMS DAs.
From the point of view of its profile, the platykurtic DA has a
broadness similar to the DSE one, being downward concave, but has
suppressed endpoint regions like the BMS DAs, thus somewhat ``merging''
these DAs, as argued in \cite{Stefanis:2014nla}.
On the other hand, the predictions obtained within our framework with
the endpoint-enhanced DSE-based DAs mostly overestimate the current TFF
data.
The distinct behavior of the presented forecasts
--- scaling vs. growth with $Q^2$ ---
for the pion TFF can serve
to select the most appropriate model DA with the help of the expected
Belle~II data in the near future.
Theoretically, it would be of great importance to explore in
detail the relation of NLCs to the DSE framework.

\begin{acknowledgements}
This work was partially supported by the Heisenberg--Landau
Program (Grant 2014), the
Russian Foundation for Fundamental Research under
Grants No.\ 12-02-00613a and No.\ 14-01-00647, and the
JINR-BelRFFR grant F14D-007.
\end{acknowledgements}

\end{document}